\documentclass[twocolumn,showpacs,preprintnumbers,amsmath,amssymb]{revtex4}

\usepackage{graphicx}
\usepackage{graphicx,epstopdf}
\usepackage{dcolumn}
\usepackage{bm}
\usepackage[usenames]{color}

\begin{document}

\title{Macroscopic and Microscopic Spectral Properties of Multilayer Brain Networks during Local and Global Synchronization }

\author{Vladimir A. Maksimenko${}^{1}$}
\author{Annika L\"{u}ttjohann${}^{2}$}
\author{Vladimir V. Makarov${}^{1}$}
\author{Mikhail V. Goremyko${}^{1}$}
\author{Alexey A. Koronovskii${}^{3}$}
\author{Anastasia E. Runnova${}^{1}$}
\author{Gilles van Luijtelaar${}^{4}$}
\author{Alexander E. Hramov${}^{1}$}
\author{Stefano Boccaletti${}^{5,6}$}
\vskip .9in
\affiliation{%
	$^{1}$Yuri Gagarin State Technical University of Saratov, REC ``Nonlinear Dynamics of Complex Systems", Saratov, 410054, Russia \\
	$^{2}$University of M\"{u}nster, Institute of Physiology I, M\"{u}nster, 48149, Germany \\
	$^{3}$Saratov State University, Faculty of Nonlinear Processes, Saratov, 410012, Russia      \\
	$^{4}$Radboud University, Donders Centre for Cognition, Nijmegen, 6525 HR , Netherlands\\
    $^{5}$CNR--Institute for complex systems, Sesto Fiorentino, 50019, Italy \\
    $^{6}$The Italian Embassy in Tel Aviv,  Tel Aviv, 68125, Israel}

\date{\today}

\begin{abstract}
	We introduce a practical and computationally not demanding technique for inferring interactions at various microscopic levels between the units of a network from the measurements and the processing of macroscopic signals.
Starting from a network model of Kuramoto phase oscillators which evolve adaptively according to homophilic and homeostatic adaptive principles, we give evidence that the increase of synchronization within groups of nodes (and the corresponding formation of synchronous clusters) causes also the defragmentation of the wavelet energy spectrum of the macroscopic signal.
Our methodology is then applied for getting a glance to the microscopic interactions occurring in a neurophysiological system, namely, in the thalamo-cortical neural network of an epileptic brain of a rat, where the group electrical activity is registered by means of multichannel EEG. We demonstrate that it is possible to infer the degree of interaction between the interconnected regions of the brain during different types of brain activities, and to estimate the regions' participation in the generation of the different levels of consciousness.
\end{abstract}


\maketitle

\section{Introduction}

The current trends in neuroscience and neurophysiology are connected with the analysis of brain networks \cite{1, 2, 3, 23} which interact with each other to perform different types of cognitive tasks, as, e.g., the formation of a memory trace \cite{4, 5}, the processing of a visual object \cite{6,7}, or  the development (on a clinical level) of pathological rhythms like epileptic seizures~\cite{8,9}.
These interactions are often quantified by means of the degree of synchrony, which can be measured both locally (i.e. within the same brain structure), or over a more global scale (i.e. in between brain structures)~\cite{10}.
While neurophysiology aims at understanding the interplay of individual neurons \cite{11, nkr}, the majority of available data (especially those acquired from human subjects) comes from non invasive tests. These tests are made, in daily practice, under the form of electro-encephalograms (EEG) or magneto-encephalograms (MEG), functional magnetic resonance imaging (fMRI), which actually measure the (electric or magnetic) group activity of large ensembles of cells.
The focal riddle for physicists and neuro-scientists consists, therefore, in disclosing the way microscopic scale neural interactions
pilot the formation of the different activities revealed (at a macroscopic scale) by EEG and MEG signals.

In a network of active elements (like neurons in the brain) one has to distinguish between the signal sensed at a microscopic scale
(the individual, electric or magnetic, activity of a neuron) and the macroscopic signal, which is instead produced by a group (or sub-network) of elements. Processes taking place at the network's \textit{microscopic} level (such as partial or complete synchronization between units, formation of clusters, etc.) affect the spectral properties of the \textit{macroscopic} signal.

Starting from a model network of Kuramoto oscillators which evolve adaptively by means of homophilic and homeostatic adaptive principles, we give evidence that the increase of synchronization within groups of nodes (leading to the formation of  structural synchronous clusters) causes also the defragmentation of the wavelet energy spectrum of the macroscopic signal, and introduce a practical and computationally not demanding technique which allows an estimation of the interaction between microscopic units of a network by means of
appropriate treatment of the macroscopic signals.
Our methodology  is then applied for getting a glance to the microscopic interactions occurring in a neurophysiological system, namely, in the thalamo-cortical neural network of an epileptic brain of a rat, where the group electrical activity is registered via multichannel EEG. We demonstrate the possibility to determine the degree of interaction between the interconnected regions of the brain during different types of brain activities.

\section{Model network of Kuramoto oscillators}

The network under consideration is a multiplex \cite{24} graph, in which topology and dynamics of nodes mutually interact via homophily \cite{18, 19} and homeostasis \cite{20} principles, as proposed in Refs.~\cite{16, 17}. The units of the network are Kuramoto oscillators~\cite{12}, the most common and simplest way to describe synchronous phenomena occurring in nature~\cite{13, 14, 22}. The system consists, therefore, of $M$ layers, each of which are made of $N=300$ oscillators. The phase evolution of each oscillator is given by
\begin{equation}
\label{eq:Kuramoto}
\dot{\phi}_i^l(t)=\omega_i+\lambda_1\sum_{j \neq i}^Nw_{ij}^l(t)\sin(\phi_j^l-\phi_i^l)+\lambda_2\sum_{j\neq l}^Msin(\phi_i^j-\phi_i^l).
\end{equation}
Here, $\phi_i^l(t)$ is the time-dependent phase of the $i$-th oscillator of  the $l$-th layer, dot denotes a temporal derivative, $\{ \omega_i \}$  is a set of randomly assigned frequencies (uniformly distributed in the interval $[1,10] \  (s^{-1})$). Notice that (in the spirit of a multi-layer network)  the natural frequency $\omega_i$ of the $i$-th oscillator is the same in all $M$ layers, whereas its instantaneous phase is actually layer-dependent. $\lambda_1$  and $\lambda_2$  are the intra- and inter- layer coupling strengths, respectively.

At first, all weights $\{ w_{ij}^l (t=0) \}$  are randomly assigned in the range $[0,1]$ (except those corresponding to $i=j$, which are all set to zero), and normalized via the condition $\sum_{j\neq i}^N w_{ij}^l=1$ ($l=1, ..., M$). The latter implies that the input strength received by each unit $i$ in each layer $l$ is constant, as in a homeostatic process.
Eq. (\ref{eq:Kuramoto}) is then initially simulated  with fixed weights $\{ w_{ij}^l(t)= w_{ij}^l (0) \}$, up to $t^A=500$~s, when the weights   start instead to evolve adaptively  (so as to enable layers to possibly reorganize) corresponding to a homophily principle, which is expressed by the following equation
\begin{equation}
\label{eq:homophily}
\dot{w}_{ij}^l(t)=p_{ij}^l(t)-\left(\sum_{k\neq i} p_{ik}^l(t)\right)w_{ij}^l(t).
\end{equation}
Here, the time dependent quantity $p_{ij}^l(t)$  is defined as
\begin{equation}
\label{eq:ppp}
p_{ij}^l(t)=\frac{1}{T}\left|\int_{t-T}^te^{\mathrm{i}(\phi_i^l(t')-\phi_j^l(t'))}dt'\right|.
\end{equation}
Notice that  $p_{ij}^l$ quantifies the average phase correlation  between oscillators $i$  and $j$ of layer $l$ over a characteristic memory time $T$. Eq. (\ref{eq:homophily}) yields the strengthening of the connections between those units which are phase correlated across $T$, verging therefore the well-known Hebbian learning process \cite{26}.

From the solutions of Eqs. (\ref{eq:Kuramoto},\ref{eq:homophily}), the {\it microscopic} signal of each unit in each layer can be estimated at all times  as $x_i^l(t)=\sin\left( \phi_i^l(t) \right)$, while the \textit{macroscopic} signal of each layer can be expressed as $X_l(t)=\sum_{i=1}^N x_i^l(t)$. The spectral properties are analyzed by the wavelet transform \cite{15}, a well-known tool suitable for analysis of various non stationary processes (as it is the present case, where signals come from layers whose structure of connections evolves actually in time).

As an initial, descriptive, example, we start with illustrating the case of a mono-layer network [$M=1$ and $\lambda_2=0$ in Eq. (\ref{eq:Kuramoto})], for an intra-layer coupling value $\lambda_1=0.5$, and for $T=100 \ s$. This parameter corresponds to the partial synchronization in the model network and the emergence of structural clusters in its topology \cite{16}.

\begin{figure}[ht]
\centering
\includegraphics[width=1.0\linewidth]{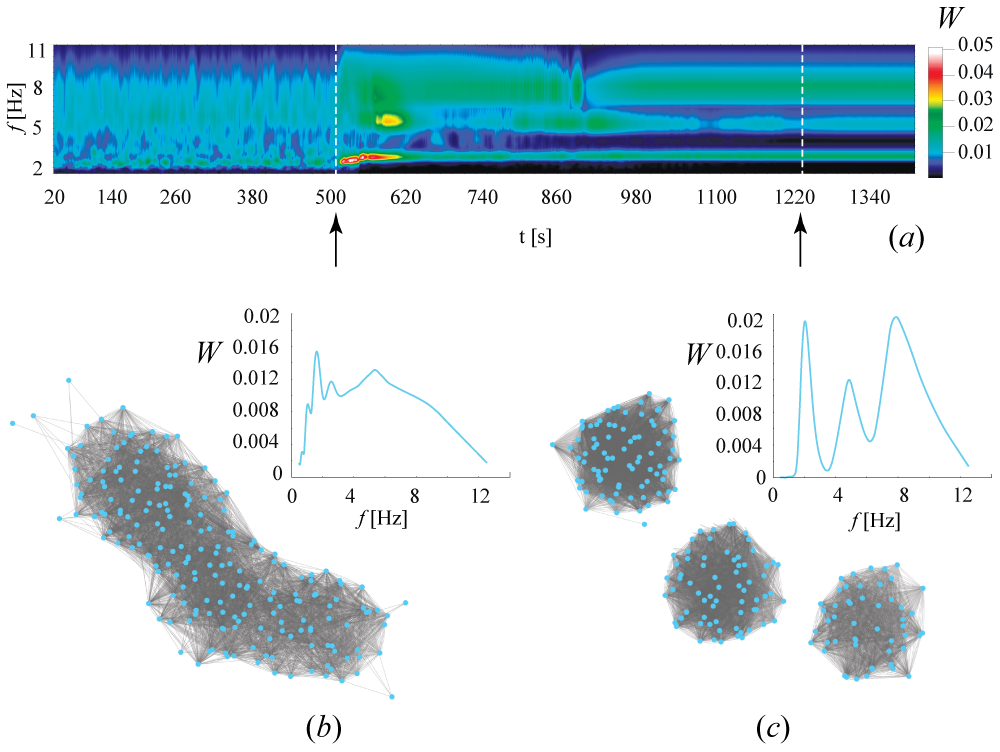}
\caption{(Color online) Panel (\textit{a}): wavelet decomposition of the macroscopic signal, produced by a monolayer network of phase oscillators under the effect of adaptive mechanism. Lower row: momentum distributions of the wavelet energy and the network topologies. Panel (\textit{b}) refers to  $t_1=500$~s (i.e. before adaptation starts to take place in the system), while panel (\textit{c}) is obtained for $t_2=500$~s (i.e. after the adaptive process produced a time independent network's topology). Both instants at which the lower panels are obtained, are indicated with vertical arrows in panel (\textit{a}). $\lambda_1=0.5$, $T=100$~s. See main text for all other specifications.
}
\label{fig:fig1}
\end{figure}

To examine the dynamics $X(t)$ in both the time and frequency domains the wavelet power spectrum can be calculated as $W(f,t)=|M(f,t)|^2$, where $M(f,t)$ is the complex wavelet surface defined with the aid of continuous wavelet transform \cite{15_, 15}
\begin{equation}\label{wavelet}
  M(f,t') = \sqrt{f}\int_{-\infty}^{+\infty}X(t)\psi^{*}\left((t-t')f\right)\,dt
\end{equation}
(the symbol ``$*$'' denotes the complex conjugation) with the Morlet mother wavelet
\begin{equation}\label{Morlet}
  \psi(\zeta)=\frac1{\sqrt[4]\pi}\exp(j2\pi\zeta)\exp\left(-\frac{\zeta^2}{2}\right).
\end{equation}

Fig.~\ref{fig:fig1},~\textit{a} reports the evolution of the wavelet power spectrum $W(f,t)$ of the macroscopic signal $X(t)$.  In the absence of the adaptive mechanisms $(t<t^A)$, the wavelet energy is distributed almost homogeneously over all range of frequencies (see the plot $W(f)$ in  Fig.~\ref{fig:fig1},~\textit{b}), as well as the network  is a single-component graph characterized by a highly homogeneous distribution in the link strengths. As $t$ exceeds $t^A$, the network structure evolves: the links between synchronized elements are strengthened, while the weakly synchronized nodes progressively loose their connections. After transient processes has expired, this leads to the appearance of three well-structured clusters, within which the elements exhibit frequency synchronization (see Fig.~\ref{fig:fig1},~\textit{c}).
The stationary momentum wavelet spectrum exhibits therefore three isolated peaks (corresponding to the spectral components of the signals taken from the three different clusters) and the structure of the layer becomes  modular (with the three frequency clusters forming three {\it densely connected} clusters of nodes, as schematically illustrated in the lower plot of Fig.~\ref{fig:fig1},~\textit{c}).

\begin{figure}[ht]
\centering
\includegraphics[width=1.0\linewidth]{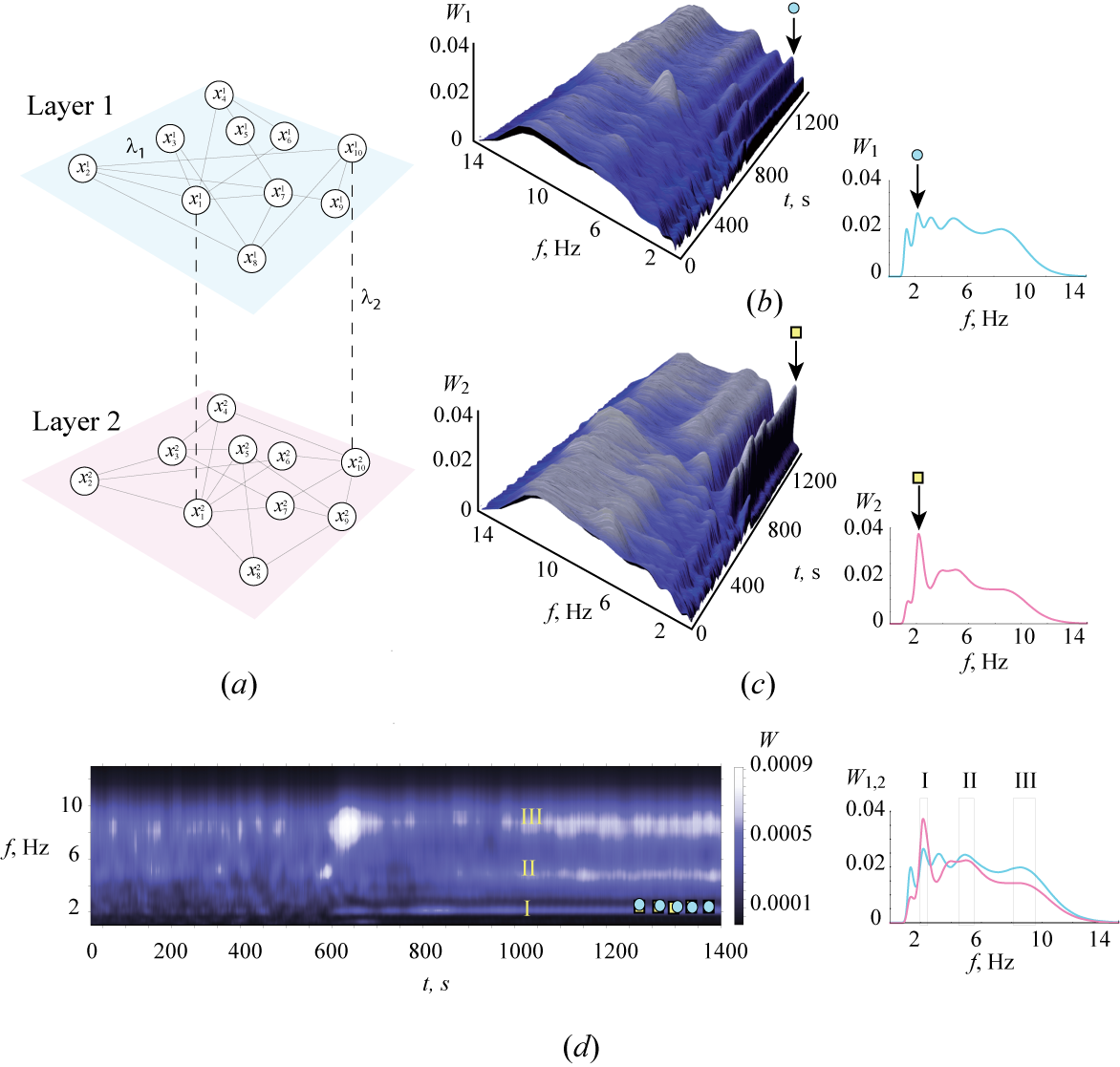}
\caption{(Color online) (\textit{a}): schematic illustration of the two-layered network.
Top right panels:  $W_1(f,t)$ [panel (\textit{b})] and $W_2(f,t)$ [ panel (\textit{c})] (see text for definitions), and the momentum distributions of the wavelet energy $W_1(f,t^*)$ [blue curve in panel (\textit{b})] and $W_2(f,t^*)$ [red curve in panel (\textit{c})], obtained for $t^*=1400$~s. (\textit{d}): the wavelet surface $W(f,t)=W_1(f,t)\times W_2(f,t)$. $\lambda_1=0.5$, $\lambda_2=0.005$, $T=100$~s. See main text for all other specifications.
}
\label{fig:fig2}
\end{figure}

A much richer scenario characterizes the case of a two-layered network \cite{16,17,18,19,20}, which is schematically illustrated in  panel (\textit{a}) of Fig.~\ref{fig:fig2}, and corresponds to setting $M=2$, $\lambda_2=0.005$, $\lambda_1=0.5$  in Eq. (\ref{eq:Kuramoto}). Here, the units have two types of connections: one set of links accounting for intra-layer interactions (i.e. those befalling among elements of the same layer), and another set of links (the inter-layer links) determining the coupling between elements of different layers. Under the simultaneous effect of intra- and inter-layer couplings, the elements belonging to different layers group into either identical and synchronous(for relatively large values of $\lambda_2$) or distinct and asynchronous  among them (for small values of $\lambda_2$) clusters. The used set of parameters corresponding to partial synchronization between the layers, that implies presence of identical clusters along with non-identical ones. The macroscopic signals produced by the two layers are $X_{1,2}(t)=\sum_{i}^N\sin(\phi_i^{1,2}(t))$.

The upper right part of Fig. \ref{fig:fig2} reports the wavelet energy distributions $W_{1,2}(f,t)$  for $X_{1}$ [panel (\textit{b})] and $X_{2}$ [panel (\textit{c})]. As in the previous case,  one clearly sees that the wavelet energy is  uniformly distributed for $t<t^A$  in both layers. For $t>t^A$, the spectral properties of each layer become similar to those of the single-layer case (the adaptive process within each layer leads to the formation of clusters and thus to a fragmentation of the wavelet spectrum) but, due to the initial mismatch in the inter-layer topology, the spectra of the two layers are different [compare the momentum distributions $W_{1,2}(f,t^*)$, for $t^*=1,400$~s, reported at the right of both panels (\textit{b}) and (\textit{c})].

While fragmentation of the wavelet spectra is caused by \textit{local} synchronization (affecting independently the oscillators within each layer), several frequency ranges exhibit a local increase of the wavelet energy in both layers. The latter is associated with the emergence of  \textit{global}  synchronization, and is a consequence of the action of intra-layer connections.
Such frequency bands are easily localizable on the surface of the wavelet spectra product $W(f,t)=W_1(f,t)\times W_2(f,t)$. In Fig.~\ref{fig:fig2},~(\textit{d}), the bands where local maxima of $W(f,t)$ are observed are marked as I, II, and III. In particular, the frequency band I is characterized by a significant increase of the wavelet energy in the both layers. Inside band I, the frequency value where the wavelet energy of the first (second) layer reaches its maximum is marked by a solid circle (square) in panel (\textit{b}) [(\textit{c})] of the same Figure. The right plot of panel (\textit{d}) discloses that the locations of the local maxima of $W_1$ and $W_2$ inside region I match perfectly, which, in turn, is a proof of synchronization between the layers.

Next, we consider the case of a 5-layered network ($M=5$), with inter- and intra-layer links arranged in a way similar to the  2-layer case.  The degree of the similarity $v_{i,j}$   between layers $i$ and $j$  can be defined (at a macroscopic scale) as the integral of the difference between the corresponding wavelet energy distributions $W_{i,j}(f,t^*)$ taken at $t^*=1400$~s (i.e. at the moment at which a stationary state has emerged, and the network topology does not change in time, except for small, residual, fluctuations affecting the intra-layer links $w_{i,j}$):
\begin{equation}
\label{eq:similarity1}
v_{i,j}=\left[\int_{f_1}^{f_2}|W_i(f)-W_j(f)|df\right]^{-1},
\end{equation}
where  $f_1=2$~Hz and $f_2=10$~Hz  are the bounds of the frequency range for which {\it global} synchronization is analyzed.

\begin{figure}[ht]
\centering
\includegraphics[width=1.0\linewidth]{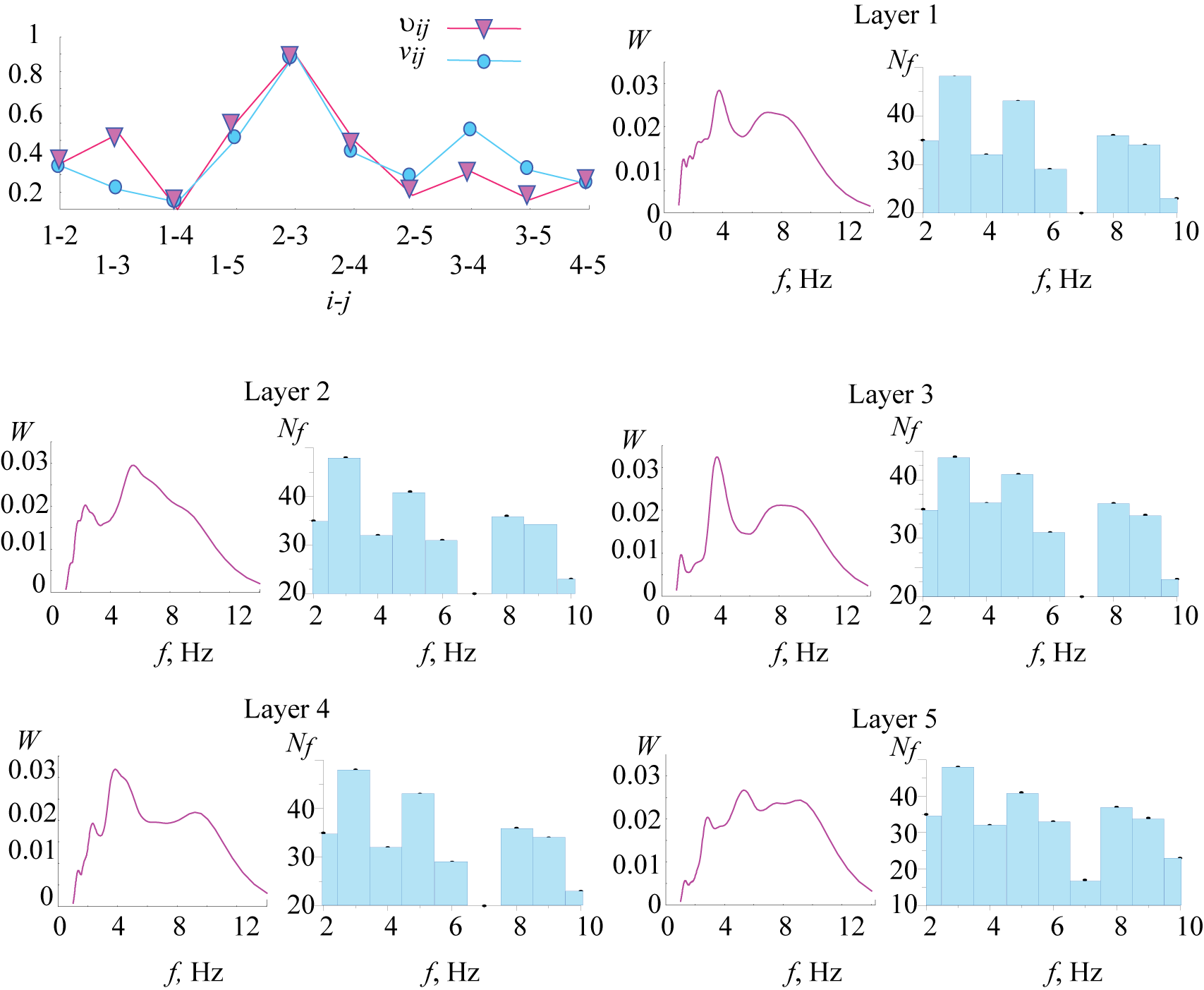}
\caption{(Color online)  Five-layered network ($M=5$) for $\lambda_1=0.5$, $\lambda_2=0.005$, $T=100$~s. The upper left panel reports the two similarity functions of Eqs. (\ref{eq:similarity1}) and (\ref{eq:similarity2}) (see text for definitions). All other panels report the wavelet energy spectra $W(f)$ (left plots) and the effective frequency distributions $N(f)$ (right histograms) for the different layers, calculated at $t^*=1500$~s.
}
\label{fig:fig3}
\end{figure}

The obtained values are shown in Fig.~\ref{fig:fig3} (upper left panel, with solid, blue, circles). One easily sees that, due to the interplay between the adaptive evolution of links within each layer and the inter-layer interactions (which may differ for different pairs of layers), the degree of similarity $v_{i,j}$ is distributed inhomogeneously [i.e. without an apparent order with respect to the pairs ($i.j$) of layers].

Likewise, one can define another degree of similarity $\nu_{i,j}$ (at a microscopic scale, i.e. based on the states of the individual oscillators) as
\begin{equation}
\label{eq:similarity2}
\nu_{i,j}=\left[\int_{f_1}^{f_2}|N_{f_i}(f)-N_{f_j}(f)|df\right]^{-1},
\end{equation}
where now  $N_{f_i}(f)$ is the effective frequency distribution of the oscillators in the i-th layer (i.e. the conveniently normalized number of oscillators of the i-th layer displaying an effective frequency $f$). The values of $\nu_{i,j}$  reflect then the similarity in terms of the number of oscillators (belonging to the i-th and j-th layers) exhibiting synchronous oscillatory behavior at the selected frequency, or equivalently, the similarity of structural clusters formed within the two layers. The upper left panel of Fig.~\ref{fig:fig3} reports also (with solid, pink, triangles) the values of $\nu_{i,j}$ for each pair of layers, and immediately beholds the remarkably good correspondence between the two measures (\ref{eq:similarity1}) and (\ref{eq:similarity2}).

The examples we made demonstrate that homophilic and homeostatic adaptive principles ordain the structure and dynamics of multi-layer networks into states where \textit{local}  and \textit{global} synchronization (within specific frequency bands) can be revealed from measurements of  microscopic and macroscopic quantities. This conclusion is of relevance for the experimental study of neural networks in the brain, where  spectral properties are actually related to the various forms of sleep-wake and cognitive activity.

\section{Neurophysiological data}
At a physiological level, available recordings (through EEG or MEG) provide typical samples of brain dynamics at its macroscopic level. They reflect, indeed, integrated extra-cellular voltage changes of neural ensembles, located in the vicinity of the recording electrode.
In parallel (and with the aid of intra- or extra-cellular single unit recordings), an operator is endowed with the opportunity of browsing on the activity of a single neuron, inspecting brain dynamics at its microscopic level.

In the current study the local field potential (LFP) and single unit recordings were obtained in 3 months-old  Genetic Absence Epilepsy Rats (GAERS), a very similar genetic absence model as the earlier mentioned WAG/Rij model,  under Fentanyl/Droperidol anaesthesia. A 1 $\mathrm{M\Omega}$ Tungsten electrode was lowed in the deep layers of the somatosensory cortex, and another 1 $\mathrm{M\Omega}$  Tungsten electrode was placed into the posterior thalamic nucleus. LFP and unit recordings of a given brain structure were gathered by the same electrode. Band-pass filters between 500 Hz and 10 kHz were applied for the unit recordings, digitalized at a constant sampling rate of 20 kHz by the SPIKE2 recording software \cite{spike2,spike22}. LFP signals were filtered between 1-100 Hz and digitalized at a constant sampling rate of 1 KHz. All experimental procedures were performed in accordance with the guidelines of the council of the European Union of the 24th November 1986 (86/609/EEC), which were approved by local authorities (review board institution: Landesamtf\"{u}rNatur, Umwelt und Verbraucherschutz Nordrhein-Westfalen; approval ID number: 87-51.04.2010.A322).

\begin{figure}[h!]
\centering
\includegraphics[width=1.0\linewidth]{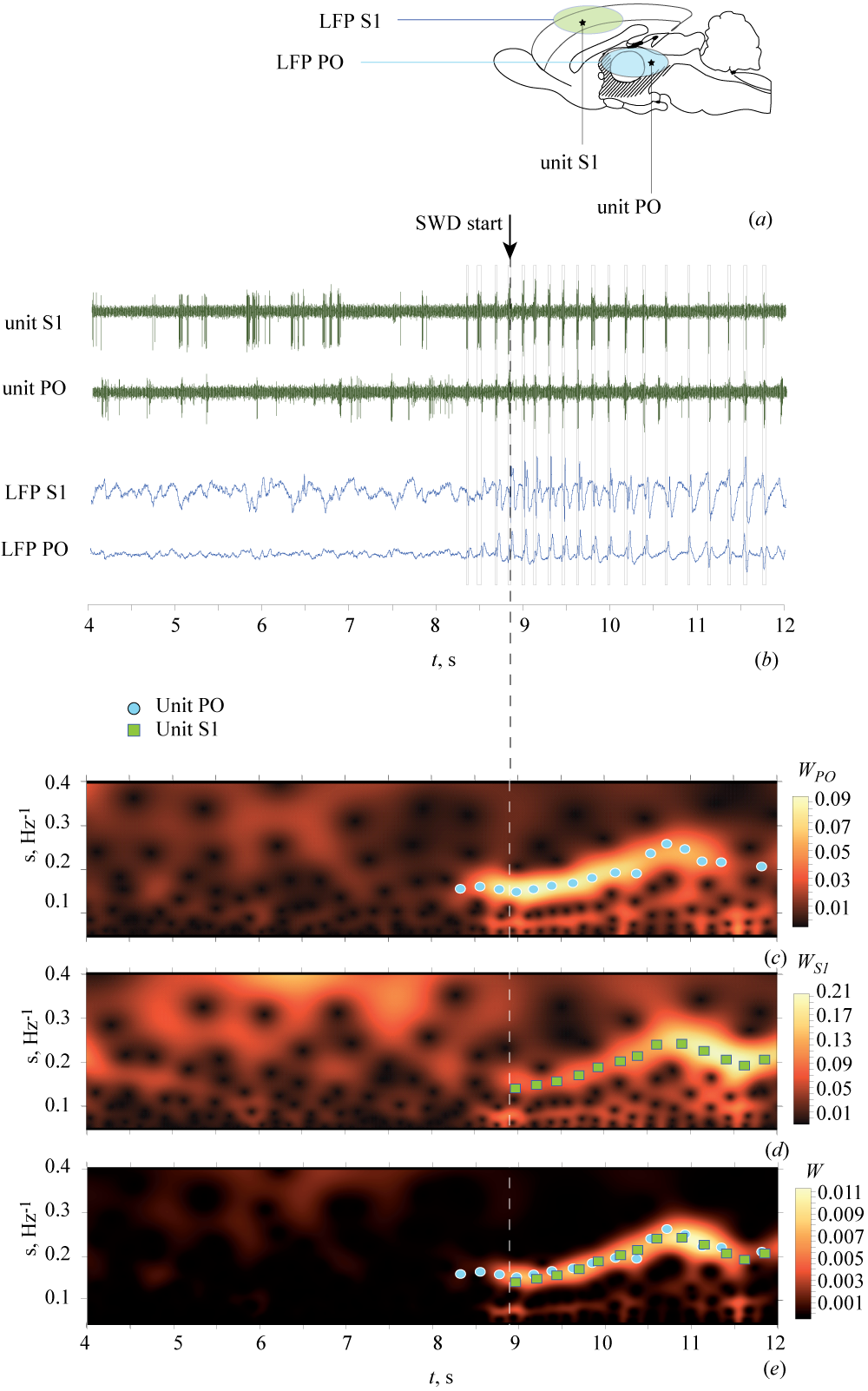}
\caption{(Color online) (\textit{a}) Schematic illustration of the experimental setup. The panel is a drawing of the rat's brain, with  marked regions of cortex and thalamus which contain the location of the recording electrodes, that actually register the group neuronal activity by means of local-field potential (LFP), and the activity of the single cells (unit recordings). (\textit{b}) The set of registered neurophysiological signals, reflecting the activity of single cells (unit S1 and unit PO) and the group activity in cortex and thalamus (LFP S1 and LFP PO). Panels (\textit{c}- \textit{e}) -- Wavelet decompositions of the macroscopic signals. (c) $W_{\mathrm{PO}}(s,t)$ obtained from signal LFP PO, (d) $W_{\mathrm{S}1}(s,t)$ obtained from signal LFP S1,  and (e) $W(s,t)=W_{\mathrm{S}1}(s,t)\times W_{\mathrm{PO}}(s,t)$. The solid circles and the solid squares shows the main spectral component of the signals, taken from the single cell in thalamus (unit PO) and cortex (unit S1), respectively. $s=1/f$ indicates the timescale, and $f$ the linear frequency. The instant of time at which there is the onset of SWD is shown by an arrow, and marked by a vertical white dashed line.
}
\label{fig:fig4}
\end{figure}
Panel (\textit{a}) of Fig. \ref{fig:fig4} illustrates the typical setup under which measurements are performed.  In panel (\textit{b}) of Fig.~\ref{fig:fig4}, we report the record of the EEG signal,  as well as that of one of its underlying microscopic components (the activity of a single neuron) measured by the same electrode as the EEG signal. Macroscopic and microscopic signals are acquired in two different (yet reciprocally connected) brain structures: cortex (S1) and thalamus (PO), which enables us to give ground to the above discussion about interrelationships between local and global synchronization processes.

In the recordings, a transition is observed from {\it normal, physiological} brain activity of the rat towards a {\it pathological, hyper-synchronous} behavior, corresponding to the occurrence of an epileptic seizure (the so-called spike-wave discharge [SWD] due to the specific waveform of the electroencephalographic signals with well-pronounced large amplitude sharp peaks and slow-waves \cite{Sitnikova:2006Coherence, Sitnikova:2009_WVLTs}), and involving an abiding change in local and global synchronization properties of the brain network.

Before seizure starts, the registered macroscopic activities of cortex and thalamus are complex signals with continuous power spectra (See Fig.~\ref{fig:fig4},~\textit{c,d,e} for $t<8.4$ s). Such a behavior actually corresponds to cells firing spontaneously and in a uncorrelated manner. At the onset of the seizure, instead, cells of cortex and thalamus start to exhibit a correlated bursting activity, which gives rise to the regularly repeated spike pattern (followed by a wave pattern) that is seen in the macroscopic recordings (Fig.~\ref{fig:fig4},~\textit{b} for   $t>8.4$~s).

Like in the case of the network model, the distributions of  wavelet energy $W_{\mathrm{S}1}(s,t)$  and  $W_{\mathrm{PO}}(s,t)$, $s=1/f$ (obtained from the macroscopic signals S1 and  PO) change from an almost homogeneous configuration ($t\leq 8.4$ s) to a shape that is characterized by a local peak positioned in the frequency band corresponding to the epileptiform activity. According to the above discussion, this fact
reveals that the onset of epileptic seizure establishes local synchronization {\it within} both cortex and thalamus, as well as global synchronization {\it between} these two regions of the brain. Indeed, by comparison of the surfaces  $W_{\mathrm{S}1}(s,t)$  and  $W_{\mathrm{PO}}(s,t)$, and  by consideration of the surface  $W(s,t)=W_{\mathrm{S}1}(s,t)  \times W_{\mathrm{PO}}(s,t)$, one can see that the considered cells in the cortex and in the thalamus are synchronized at the frequency of the seizure and, moreover, they are synchronized with other cells belonging to the same part of the brain (See Fig.~\ref{fig:fig4},~\textit{e} for   $t>8.4$~s).

Following the approach described in the previous section, one can analyze the dynamics of the thalamo-cortical network by means of a multichannel set of EEG recordings taken from  Wistar Albino Glaxo from Rijswijk (WAG/Rij) rats  \cite{27} --- a genetic animal model giving rise to absence epilepsy. In the experiment 6 month old WAG/Rij rats were chronically implanted with stainless steel electrodes in layer 4 to 6 of the somatosensory cortex, as well as in {\it i)} the posterior thalamic nucleus, {\it ii)} the ventral-postero-medial thalamic nucleus, {\it iii)} the anterior thalamic nucleus, and {\it iv)} the reticular thalamic nucleus under deep isoflorane anaesthesia. Two weeks after surgery, EEG signals were recorded from theses structures in freely moving animals. Signals were filtered by a band pass filter with cut-off points at 1(HP) and 100(LP) and a 50 Hz Notch filter and digitalized by WINDAQ-recording-system (DATAQ-Instruments Inc., Akron, OH, USA) \cite{Windq,Windq2} with a constant sample rate of 500 Hz. Experiments were  carried out in accordance with the Ethical Committee on Animal Experimentation of Radboud University Nijmegen (RU-DEC).

As the result, the recordings taken from three cortical and four thalamical electrodes  are considered at different instants of time: {\it i)} at the beginning of a seizure, {\it ii)} at the end of a seizure, {\it iii)} at a time at which the rat is in a state of active wakefulness, and {\it iv)} at a time at which the rat is in a state of slow-wave sleep, refer to Fig.~\ref{fig:fig5}. Moreover (and according to the neurophysiological background of absence epilepsy \cite{21}), the dynamics of the network is studied within three different  frequency bands: {\it i)} $\Delta_{f_1}$  (the low-frequency, LF, oscillation range 1--5 Hz), {\it ii)} $\Delta_{f_2}$  (the range 5--10 Hz, SWD, of seizure activity), and {\it iii)}  $\Delta_{f_3}$  (the range of high-frequency, HF, theta activity, 7--20 Hz). The wavelet energies $W_{LF}$ , $W_{SWD}$  and  $W_{HF}$ are calculated as
\begin{equation}
\label{eq:eeg_energy}
W_{LF,SWD,HF}=\int_{t-\tau}^{t}\left[\int_{f\in \Delta_{f_{1,2,3}}}W_i(f,\xi)df\right]d\xi
\end{equation}
where $\tau=2.5$~s is a time lapse chosen conveniently (and judiciously) to neglect spurious fluctuations, and $i$ is the number of the EEG channel.

\begin{figure}[ht]
\centering
\includegraphics[width=1\linewidth]{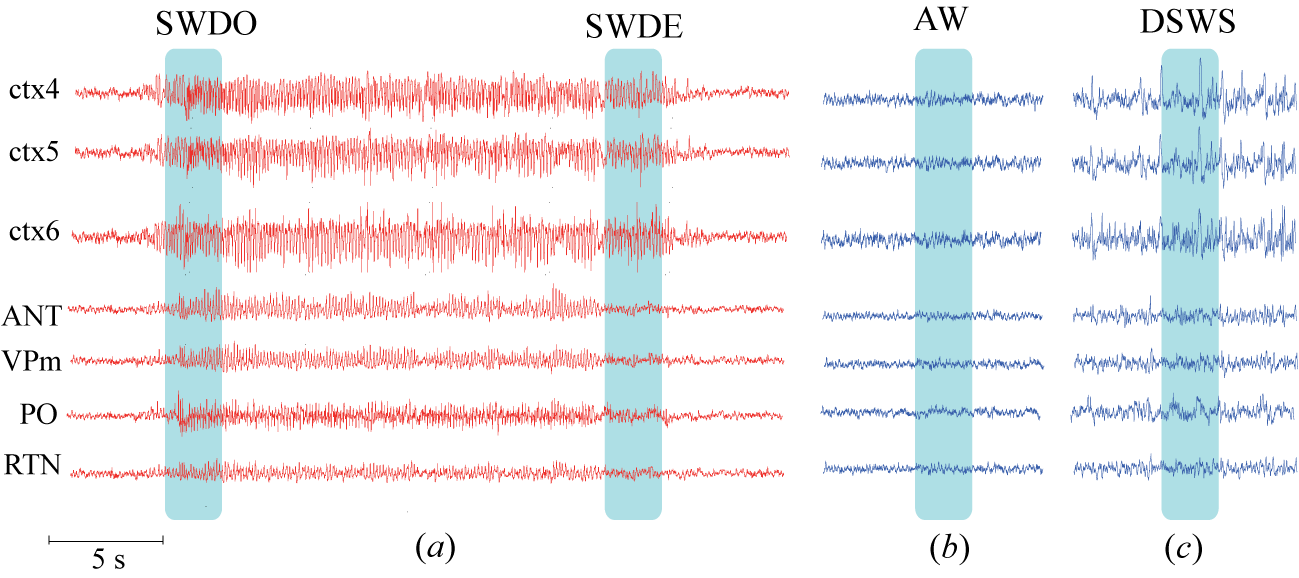}
\caption{(Color online). Typical fragments of registered EEG signals, reflecting the activity of the neuronal group
in cortex and thalamus during the epileptic discharge (SWD, a), during active wakefulness (AW,b), and during deep slow-wave sleep (DSWS, c). The epochs
corresponding to the onset (SWDO) and end (SWDE) of SWD, to AW and DSWS are marked by rectangles.
}
\label{fig:figExtras}
\end{figure}

\begin{figure}[ht]
\centering
\includegraphics[width=1.0\linewidth]{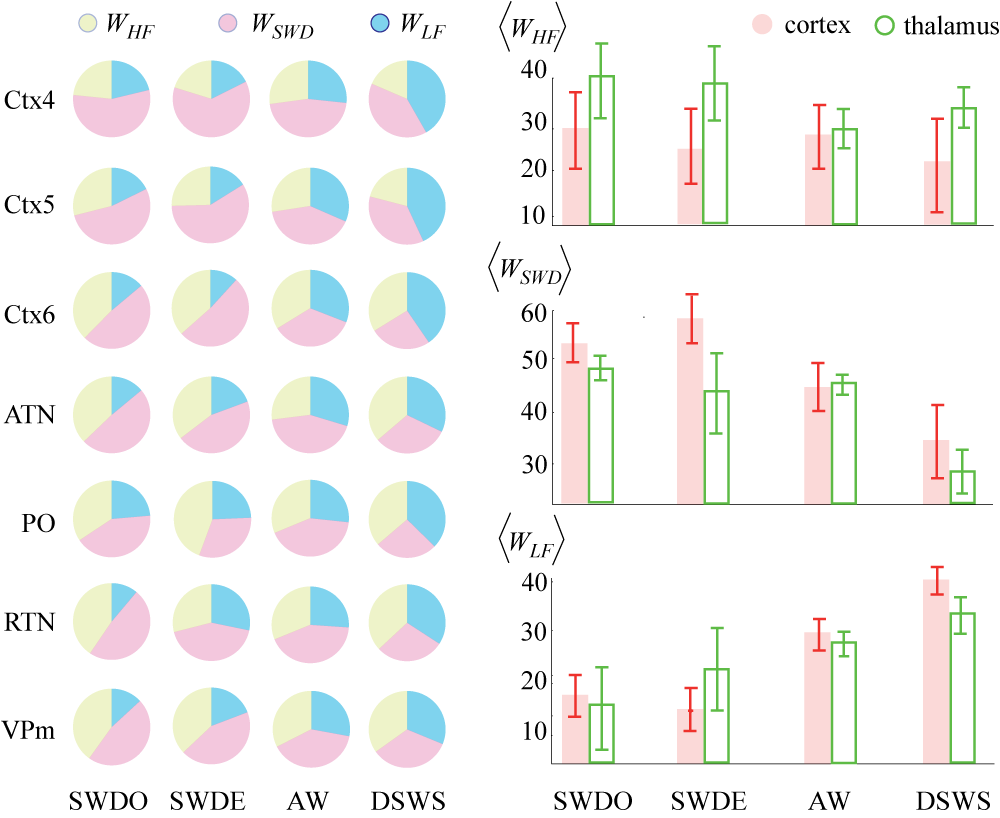}
\caption{(Color online). Left panel: the ratio of the wavelet energies [Eq. (\ref{eq:eeg_energy})], calculated for all considered states (columns) and all channels (rows). Right panel: the energy values [Eq. (\ref{eq:eeg_energy})] averaged over the cortical (solid bars) and thalamical (empty bars) channels. The error bars stand for the maximal deviation within the channel set. Once again, SWDO, SWDE, AW, and  DSWS stay for the SWD onset, the SWD end,  the active wakefulness,  and the deep slow-wave sleep, respectively.
}
\label{fig:fig5}
\end{figure}

The characteristic dynamical states of cortico-thalamo-cortical neuronal network of epileptic brain  [the SWD (at the onset and the end), active wakefulness, and deep slow-wave sleep] are illustrated in Fig.~\ref{fig:figExtras}, which reports typical EEG fragments recorded in cortex and thalamus, with the analyzed time epoches that are marked by rectangles and labeled as SWDO, SWDE, AW, and DSWS, respectively.
Fig.~\ref{fig:fig5} (left panel) illustrates the ratio of the wavelet energies [Eq. (\ref{eq:eeg_energy})] calculated for all considered states (columns) and all channels (rows). One easily sees that during SWDO (i.e. when the epileptic seizure has just recently started), most of the wavelet energy (about $50\%$) is concentrated in the range of SWD activity ($W_{SWD}$) both in the cortex and in the thalamus. This reflects the fact that the synchronization level between neurons in the thalamo-cortical network of the brain increases at those frequencies which correspond to the band of the epileptic seizure $\Delta_{f_2}$. As the end of the seizure is approached (during SWDE) synchronization in the frequency band of the epileptic seizure $\Delta_{f_2}$ increases in the cortex, leading to an increase of $W_{SWD}$.  At the same time, neurons in the thalamus start to go out of the hyper-synchronous state, and the value of $W_{SWD}$  decreases in the thalamus recordings.

During the normal physiological activity, i.e. active wakefulness (AW) state, the value of $W_{SWD}$ is less in the cortex. At the same time, the wavelet energy increases in the frequency bands $\Delta_{f_1}$ and $\Delta_{f_3}$ both in the cortex and in the thalamus, which is caused by the neurons being involved in other types of brain activities (as, for instance cognitive tasks). During deep slow-wave sleep (DSWS), most of the wavelet energy is concentrated in the low frequency band $\Delta_{f_3}$, and $W_{LF}$ experiences a significant growth, which actually corresponds to an increased number of the neurons producing low-frequency rhythms, and global synchronization between them both in the cortex and in the thalamus.

In order to quantitatively compare the changes in wavelet energies for the different states of the epileptic brain we further report the values  $\langle W_{LF}\rangle$, $\langle W_{SWD}\rangle$  and $\langle W_{HF}\rangle $, which are the energy values [following Eq. (\ref{eq:eeg_energy}] averaged over the cortical/thalamical recordings. These averaged values are shown in the right panel of Fig.~\ref{fig:fig5}, where solid and empty bars correspond to the cortex and thalamus recordings, respectively. The obtained results show that during the transition from the seizure onset to the seizure end (SWDO$\,\rightarrow \,$SWDE)  neurons in the cortex exhibit a growth of the synchronization level in the SWD frequency band $\Delta_{f_2}$, while the synchronization level in the thalamus decreases. As a consequence, the fact that the thalamic neurons abandon the synchronous state, seems to be the cause of the epileptic discharge destruction.

With the increase of $\langle W_{SWD}\rangle$, the values of $\langle W_{LF}\rangle$ and $\langle W_{HF}\rangle$  decrease in the cortex, while the vice-versa occurs in the thalamus. $\langle W_{SWD}\rangle$  decreases in the SWD frequency band with the growth of $\langle W_{LF}\rangle$, whereas $\langle W_{HF}\rangle$ remains practically unchanged. Such changes expose that the neuron groups in the cortex and thalamus interact with different intensities in the different frequency bands, depending on the specific brain state. For instance,  synchronization in the SWD frequency band in the cortex becomes stronger at the end of the epileptic discharge. At the same time, in the thalamus one observes a decrease of synchronization in the SWD frequency band, and and increase in low- and high-frequency bands, i.e. the neurons start to interact more intensively producing the low- and high-frequency activity. During  AW  and DSWD, one observes a significant increase of the wavelet energy $\langle W_{LF}\rangle$   with a simultaneous decrease of the energy, corresponding to the epileptiform pattern $\langle W_{SWD}\rangle$ and alpha/theta activity $\langle W_{HF}\rangle$.

A first conclusion that can be drawn is  that  neurons in the cortex and thalamus interact in the various frequency bands with  different intensities depending on the specific state, leading to a variety of synchronous patterns. It should be noted that, along with the interaction between neurons located relatively close to each other in the cortex layers, the interaction between remote regions of the cortico-thalamo-cortical network, e.g. different nuclei of the thalamus, is also relevant for understanding processes leading to the hyper-synchronous epileptic dynamics.

{Like in the case of the multilayered network model, one can analyze the degree of the interaction between the different parts of the cortico-thalamo-cortical network during the generation of the different forms of activity. It can be assumed that in brain the neurons belonging to the different brain structures can be involved together in the generation of the certain rhythm. In this case the wavelet spectrums of the EEG signals, taken from these areas of brain, are expected to demonstrate the increase of the similarity. According to this, the strength of the interaction between the corresponded areas of brain can be estimated by the calculation of the degree of similarity $w_{i,j}$  via Eq.~(\ref{eq:similarity1}), where the limits of integration are chosen accordingly to the frequency bands associated with the type of the rhythm (in our study the limits are defined by the bands $\Delta_{f_{1,2,3}}$).}

\begin{figure}[ht]
\centering
\includegraphics[width=1.0\linewidth]{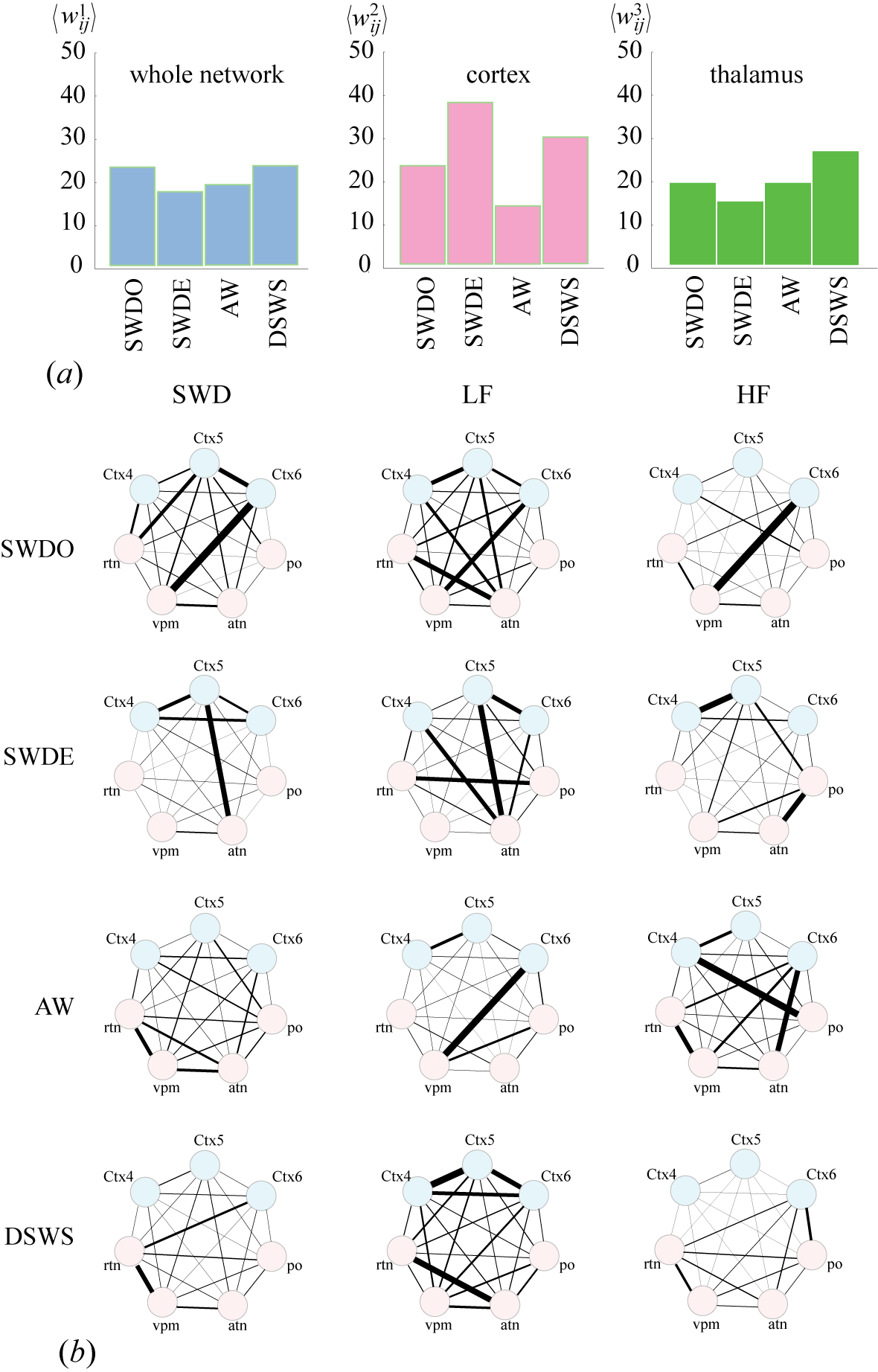}
\caption{(Color online).(\textit{a}) The mean value of the strength of interaction between the different parts of the thalamo-cortical neuronal  network. The values [calculated for whole network ($w^1_{ij}$), for the cortex ($w^2_{ij}$), and for the thalamus ($w^3_{ij}$)] are averages of the coefficients $w_{i,j}$ over the considered parts of the brain.  $w_{i,j}$ are estimated via Eq.~(\ref{eq:similarity1}), based on the similarity of the wavelet spectrums. (\textit{b}) The schematic illustration of the coefficients $w_{i,j}$, reflecting the degree of interaction between the different parts of the cortex and thalamus, for all considered states of brain network and frequency bands. The values of $w_{i,j}$ are shown by the increase (or decrease) of the line width which connect the corresponded brain structures. Other stipulations as in the Caption of Fig. \ref{fig:fig5}.
}
\label{fig:fig6}
\end{figure}

{Fig.~\ref{fig:fig6},~\textit{a} reports the mean degree of the interaction between the areas of brain belonging to the cortex, the thalamus, and the whole neuronal network, calculated by averaging the coefficients $w^k_{i,j}$ over the corresponding region of the brain network (with $k=1$ for the whole cortico-thalamo-cortical network, $k=2$ for cortex and $k=3$ for thalamus). In Fig.~\ref{fig:fig6},~\textit{b}, the  values $w_{i,j}$ are shown by the increase (or decrease) of the line width, which connect the corresponding brain structures.  From Fig.~\ref{fig:fig6},~\textit{a} one easily sees that the cortico-thalamo-cortical network is characterized by a high degree of global interaction at the onset of the seizure (SWDO state) and during the DSWS state, while the minimal value of $w^1_{i,j}$ is achieved for the SWDE state. Looking at the $w^2_{i,j}$ and $w^3_{i,j}$ values (corresponding to cortex and thalamus, respectively), one finds that the thalamical network exhibits a high level of global interaction during the SWDO and DSWD states, while during the SWDE state the different regions of thalamus interact weakly with each other. On the contrary, the cortical regions interact more strongly at the end of the epileptic seizure.}

{From Fig.~\ref{fig:fig6},~\textit{a} one can conclude that, along with the interaction within the differen parts of cortex and thalamus, the increase of the interaction {\it between} these parts manifests itself as a key feature of the epileptic seizure. Indeed, at the beginning of the SWD (i.e. when the seizure occurs spontaneously) there is an increase not only within the cortex and thalamus separately, but also between them. At the same time, at the end of seizure, despite the strong and stable cooperation between the parts of the cortex, the seizure ends due to the destruction of the interaction between cortex and thalamus.}

{From Fig.~\ref{fig:fig6},~\textit{b}, where the interaction in the cortico-thalamo-cortical network is shown for the different frequency bands, one can conclude that at seizure onset (the SWDO state)  the different parts of neuronal network demonstrate a high level of interaction in the  bands $\Delta_{f_1}$  and  $\Delta_{f_2}$ (the low- and spike-wave oscillations frequency). The increase of the interaction in the $\Delta_{f_1}$ band is related to the presence of the low-frequency delta precursors, as shown in Ref.~\cite{25}. At the spike-wave discharge end (the SWDE state),  there are still strong interactions  in the cortex, related to the SWD-frequency band. At the same time, the parts of the thalamus interact in a much weaker way in this frequency band and, moreover, exhibit a significant decrease in the interaction with cortical neurons. When the seizure is finished and the animal exhibits active wakefulness, the different parts of cortico-thalamo-cortical network start to interact stronger in the frequency band $\Delta_{f_3}$, which corresponds to the generation of high-frequency brain rhythms. During the deep slow-wave sleep, such interactions can be observed in the band $\Delta_{f_1}$  of low-frequencies, while in the other bands the different parts of the brain interact more weakly. During the AW state, the parts of the cortex and thalamus interact almost equally with each other, but the high degree of the interaction here is revealed in the high-frequencies range. The conclusion is that, during the slow-wave sleep, the low-frequency oscillations (delta-waves) are generated by the neuron populations both in the cortex and in the thalamus, and this type of brain activity is characterized by a high degree of inter-layer interaction in these parts of the brain as well as over the whole neuronal network of  cortex and thalamus.}

\section{Conclusion}
Microscopic processes characterizing the interaction between the units of a network affect the properties of the network at any macroscopic scale.
However, how this interdependency can be used in the reverse way (i.e. to reveal the nature of microscopic interactions by the study of the global changes of the network) is not generically obvious, especially when the dynamics of the single nodes is unknown. This is an important issue in neurophysiology, where data on the brain dynamics is available by means of electroencephalograms (EEG) or magnetoencephalograms (MEG), which actually measure the electric group activity of large ensembles of cells, while the state of individual neurons and the evolution of the links between them cannot be easily revealed. We have here demonstrated how to use the macroscopic properties of the network to get information on the network dynamics at a microscopic level.

Our analysis started with a model network of Kuramoto oscillators, for which the evolution of the links between the nodes is controlled by homophilic and homeostatic adaptive principles.  We gave evidence that the increase of synchronization within groups of nodes (leading to the formation of  structural synchronous clusters) causes also the defragmentation of the wavelet energy spectrum of the macroscopic signal.
Considering a multi-layer network model we revealed that nodes belonging to different layers interact with each other with  different degrees of intensity. Namely, a strong interaction between the layers reflects the appearance of  structural clusters (with the same spectral properties) on the layers. The degree of similarity between pairs of layers can be estimated as the integral of the difference between the wavelet energy distributions, and they display inhomogeneities over the considered pairs of layers.

The same phenomena were observed in a neurophysiological system considered, namely, in the cortico-thalamo-cortical network of the brain of a genetically epileptic rat \cite{28}, where the group electrical activity are registered by means of multichannel EEG. We demonstrated the possibility to determine the degree of interaction between the interconnected regions of the brain. Specifically, depending on the type of brain activity, we found that the neurons in cortex and thalamus interact in the different frequency bands with different degrees of intensity, which, in turn, leads to the formation of different synchronous patterns. Along with the interaction between neurons located relatively close to each other in the cortex layers, we gave evidence of interaction between more remote regions of cortico-thalamo-cortical network, e.g. different thalamic nuclei.

In addition, we detected strong synchronization of the cortical layers at the end of the epileptic seizures together with a decrease in their synchrony with thalamic nuclei. This is an indication of the attempt of the cortex, the location of the epileptic onset zone in this epilepsy model, to keep the seizure ongoing, which is corrupted by the thalamus.Interestingly, network analyzes of the multichannel EEG in the same epilepsy model showed a high coherence and phase consistency between cortex and thalamus and between cortical layers during the seizure \cite{29,30} in agreement to what is well known in absence epileptic patients. A recent proposed scenario on how absence seizures spontaneously end, based on different types of advanced signal analyzes, also mentioned that intra-thalamic processes heavily contribute to the spontaneous ending of the seizure \cite{31}.

Our approach constitutes a practical technique for the investigation of brain neuronal network interactions, with the potential of getting a glance at interactions at a microscopic level by the analysis of the macroscopic signals commonly acquired in neuroscience studies. In particular, our study may be applied in a wide range of neurophysiological studies, which  investigate functional brain network conductivities by means of EEG and MEG data during different forms of cognitive and behavioural tasks as well as for the study of pathophysiological brain processes.

\section{acknowledgments}
This work has been supported by the Ministry of Education and Science of Russian Federation (Projects 3.861.2017/PCH and 3.4593.2017/VY) and Russian Foundation for Basic Research (Grant 16--32--00334).

{}


\begin{thebibliography}{}


\bibitem{1} R. F. Betzel, S. Gu, J. D. Medaglia, F.Pasqualetti, and D. S. Bassett, \textit{Sci. Rep.} {\bf 6}, 30770 (2016)

\bibitem{2} A. M. Hermundstad, D. S. Bassett, K. S. Brown, E. M. Aminoff, D. Clewett, S. Freeman, A. Frithsen, A. Johnson, C. M. Tipper, M. B. Miller, S. T. Grafton, J. M. Carlson, \textit{Proceedings of the National Academy of Sciences} {\bf 110}, 6169–6174 (2013)

\bibitem{3} S. Atasoy, I.  Donnelly, and J. Pearson, \textit{Nat. Commun.} {\bf 7}, 10340 (2016)

\bibitem{23} S. Boccaletti,  V. Latora,  Y. Moreno, M. Chavez, and D.-U. Hwang, \textit{Phys. Rep.} {\bf 424}, 175 (2006).

\bibitem{4} G. Buzs\'aki, \textit{Neuroscience} {\bf 31}, 551(1989).

\bibitem{5} C. Haenschel,D. J. Vernon, P.  Dwivedi, J. H.  Gruzelier, and T. Baldeweg, \textit{J Neurosci.} {\bf 25}, 10494 (2005).

\bibitem{6} R. M. Cichy, A. Khosla, D. Pantazis, A. Torralba, and A.  Oliva, \textit{Sci. Rep.} {\bf 6}, 27755 (2016).

\bibitem{7} T. J. Palmeri, and  I. Gauthier, \textit{Nat. Rev. Neurosci.} {\bf 5}, 291 (2004).

\bibitem{8} A. Cavanna, and F. Monaco, \textit{Nat. Rev. Neurol.} {\bf 5}, 267 (2009).

\bibitem{9} H. K. Meeren, J. P. Pijn, E. L. Van Luijtelaar, A. M. Coenen, and F. H. Lopes da Silva, \textit{J. Neurosci.} {\bf 22}, 1480 (2002).

\bibitem{10} M. Jalili, \textit{Sci. Rep.} {\bf 6}, 29780(2016).

\bibitem{11} E. M. Maynard, N. G. Hatsopoulos, C. L. Ojakangas, B. D. Acuna, J. N. Sanes, R. A. Normann, and J. P. Donoghue, \textit{J. Neurosci.} {\bf 19}, 8083  (1999).

\bibitem{nkr} V. I. Nekorkin,\textit{Physics-Uspekhi} {\bf 51}, 295 (2008).

\bibitem{24}S. Boccaletti, G. Bianconi, R. Criado, C.I. del Genio, J. G\'{o}mez-Garde\~{n}es, M. Romance, I. Sendi\~{n}a-Nadal, Z. Wang, and M. Zanin, \textit{Phys. Rep.} {\bf 544}, 1-122 (2014).

\bibitem{18} D. O. Hebb, \textit{The Organization of Behavior},(Wiley, New York, 1949).

\bibitem{19} M. McPherson, L. Smith-Lovin, and J. M. Cook, \textit{Annu. Rev. Sociol.} {\bf 27}, 415 (2001).

\bibitem{20} R. I. M. Dunbar, \textit{J. Human Evo.} {\bf 22}, 469 (1992).

\bibitem{16} S. Assenza, R. Gutierrez, J. Gomez-Gardenes, V. Latora, and S. Boccaletti, \textit{Sci. Rep.} {\bf 1}, 99 (2011).

\bibitem{15_}
A. N. Pavlov, A. E. Hramov, A. A. Koronovskii, E. Yu. Sitnikova, V. A. Makarov, A. A. Ovchinnikov,  \textit{Physics-Uspekhi} {\bf55}  845 (2012),


\bibitem{15} A. Hramov, A. Koronovskii, V. Makarov, A. Pavlov, and E. Sitnikova, \textit{Wavelets in Neuroscience}, (Springer Heidelberg New York Dordrecht London, 2015).


\bibitem{17} V. V. Makarova, A. A. Koronovskiia,  V. A. Maksimenkoa,  A. E. Hramovb,  O. I. Moskalenko,  J. M. Bulduc, and S. Boccalettie,\textit{Chaos, Solitons and Fractals} {\bf 84}, 23 (2016).

\bibitem{12} Y.Kuramoto, \textit{Chemical Oscillations, Waves, and Turbulence}, (Springer-Verlag, New York, 1984).

\bibitem{13} S. H. Strogatz, \textit{Physica D} {\bf 143}, 1 (2000).

\bibitem{14} A. Pikovsky,  M. Rosenblum, and J. Kurths, \textit{Synchronization - A Universal Concept in Nonlinear Sciences}, (Cambridge University Press, 2001).

\bibitem{22} S. Boccaletti, J. Kurths,G. Osipov, D. L. Valladares, and C. S. Zhou, \textit{Phys. Rep.} {\bf 366}, 1 (2002).

\bibitem{spike2} \textit{http://ced.co.uk/products/spkovin}

\bibitem{spike22} T.Seidenbecher, R. Staak, and H. C. Pape, \textit{Eur. J. Neurosci.} {\bf 10}, 1103 (1998).

\bibitem{26} S. Song, K. D. Miller, and L. F.  Abbott, \textit{Nat. Neurosci.} {\bf 3}, 919 (2000).

\bibitem{Sitnikova:2006Coherence} E. Sitnikova, and E. L. van Luijtelaar, \textit{Epilepsy Res.} {\bf 71}, 159 (2006).

\bibitem{Sitnikova:2009_WVLTs} E. Sitnikova, A. E. Hramov, A. A. Koronovskii, and E. L. van Luijtelaar,\textit{J. Neurosci. Methods.} {\bf 180}, 304 (2009).

\bibitem{27} K. Sarkisova, and G.  van Luijtelaar,\textit{Prog. Neuropsychopharmacol. Biol. Psychiatry.} {\bf 35}, 854 (2011).

\bibitem{Windq} \textit{ http://www.dataq.com/}

\bibitem{Windq2} A. L\"{u}ttjohann, and G. van Luijtelaar, \textit{Neurobiol Dis.} {\bf 47}, 1, 49 (2012).

\bibitem{21} J. R. Tenney, H. Fujiwara, P. S. Horn, J. Vannest, J. Xiang, T. A. Glauser, and D. F. Rose, \textit{Ann. Neurol.} {\bf 76}, 558 (2014).

\bibitem{25} G. van Luijtelaar, A. Hramov, E. Sitnikova, and A. Koronovskii, \textit{Clin. Neurophysiol.} {\bf 122}, 687 (2011).

\bibitem{28} G. van Luijtelaar, and E. Sitnikova, \textit{Neurosci. Biobehav. Rev.} {\bf 30}, 983 (2006).

\bibitem{29} E. Sitnikova, and G. van Luijtelaar, \textit{Epilepsy Res.} {\bf 84}, 159 (2009).

\bibitem{30} A. L\"{u}ttjohann, J. M. Schoffelen, and G. van Luijtelaar, \textit{Exp. Neurol.} {\bf 239}, 235 (2013).

\bibitem{31} A. L\"{u}ttjohann, and E. L.  van Luijtelaar, \textit{Front. Physiol.} {\bf 6}, 16 (2015).


	
\end{thebibliography}
\end{document}